\documentclass[preprint,nonatbib,10pt,nocopyrightspace]{sigplanconf}

\usepackage{etex}

\usepackage[utf8]{inputenc}

\usepackage{csquotes}

\usepackage{booktabs}

\usepackage{tabularx}

\usepackage{array}
\newcolumntype{L}[1]{>{\raggedright\let\newline\\\arraybackslash\hspace{0pt}}m{#1}}
\newcolumntype{C}[1]{>{\centering\let\newline\\\arraybackslash\hspace{0pt}}m{#1}}
\newcolumntype{R}[1]{>{\raggedleft\let\newline\\\arraybackslash\hspace{0pt}}m{#1}}

\makeatletter

\usepackage{color}

\usepackage[%
    backend=bibtex,
    style=numeric-comp,
    sorting=none,        % nty,nyt,nyvt,anyt,anyvt,ynt,ydnt,none
    sortcites=true,      % sort \cite{b a d c}: true,false
    block=none,          % space between blocks: none,space,par,nbpar,ragged
    indexing=false,      % indexing options: true,false,cite,bib
    citereset=none,      % don't reset cites
    isbn=false,          % print ISBN?
    url=true,            % print URL?
    doi=false,           % print DOI?
    natbib=true,         % natbib compatability
    maxbibnames=99       % no 'et-al' in the bibliography pls
  ]{biblatex}

\usepackage[T1]{fontenc}
\usepackage{lmodern}
\usepackage{textcomp}

\usepackage{algorithm}
\usepackage{algpseudocode}

\algblockdefx[Loop]{Loop}{EndLoop}[1][]{\textbf{Loop} #1}{\textbf{End
    Loop}}

\algrenewcommand\ALG@beginalgorithmic{\footnotesize}

\usepackage{amsmath}

\usepackage{esvect}

\usepackage{amsfonts}
\usepackage{amssymb}

\usepackage{graphicx}
\usepackage{mathtools}

\usepackage{bm}

\usepackage{subcaption}
\expandafter\def\csname ver@subfig.sty\endcsname{}

{\endalign}

\DeclareMathOperator*{\argmin}{arg\,min}
\DeclareMathOperator*{\argmax}{arg\,max}

\newcommand*\wrapletters[1]{\wr@pletters#1\@nil}
\def\wr@pletters#1#2\@nil{#1\allowbreak\if&#2&\else\wr@pletters#2\@nil\fi}

\usepackage{gensymb}

\begin{document}

\setlength{\pdfpageheight}{\paperheight}
\setlength{\pdfpagewidth}{\paperwidth}

\conferenceinfo{ADAPT '16}{Month d--d, 20yy, City, ST, Country}
\copyrightyear{2016}
\copyrightdata{978-1-nnnn-nnnn-n/yy/mm}
\doi{nnnnnnn.nnnnnnn}

% Uncomment one of the following two, if you are not going for the
% traditional copyright transfer agreement.

%\exclusivelicense                % ACM gets exclusive license to publish,
                                  % you retain copyright

%\permissiontopublish             % ACM gets nonexclusive license to publish
                                  % (paid open-access papers,
                                  % short abstracts)

% \titlebanner{banner above paper title}        % These are ignored unless
% \preprintfooter{}   % 'preprint' option specified.

\title{Autotuning OpenCL Workgroup Size for Stencil Patterns}

% \subtitle{Subtitle Text, if any}

\authorinfo{Chris Cummins\and Pavlos Petoumenos \and Michel Steuwer \and Hugh Leather}
           {University of Edinburgh}
           {c.cummins@ed.ac.uk, ppetoume@inf.ed.ac.uk, michel.steuwer@ed.ac.uk, hleather@inf.ed.ac.uk}

\maketitle

\begin{abstract}
  Selecting an appropriate workgroup size is critical for the
  performance of OpenCL kernels, and requires knowledge of the
  underlying hardware, the data being operated on, and the
  implementation of the kernel. This makes portable performance of
  OpenCL programs a challenging goal, since simple heuristics and
  statically chosen values fail to exploit the available
  performance. To address this, we propose the use of machine
  learning-enabled autotuning to automatically predict workgroup sizes
  for stencil patterns on CPUs and multi-GPUs.

  We present three methodologies for predicting workgroup sizes. The
  first, using classifiers to select the optimal workgroup size. The
  second and third proposed methodologies employ the novel use of
  regressors for performing classification by predicting the runtime
  of kernels and the relative performance of different workgroup
  sizes, respectively. We evaluate the effectiveness of each technique
  in an empirical study of 429 combinations of architecture, kernel,
  and dataset, comparing an average of 629 different workgroup sizes
  for each. We find that autotuning provides a median $3.79\times$
  speedup over the best possible fixed workgroup size, achieving 94\%
  of the maximum performance.
\end{abstract}

% \category{CR-number}{subcategory}{third-level}

% % general terms are not compulsory anymore,
% % you may leave them out
% \terms
% term1, term2

% \keywords
% keyword1, keyword2

\section{Introduction}\label{sec:introduction}

Stencil codes have a variety of computationally demanding uses from
fluid dynamics to quantum mechanics. Efficient, tuned stencil
implementations are highly sought after, with early work in 2003 by
Bolz et al.\ demonstrating the capability of GPUs for massively
parallel stencil operations~\cite{Bolz2003}. Since then, the
introduction of the OpenCL standard has introduced greater
programmability of heterogeneous devices by providing a
vendor-independent layer of abstraction for data parallel programming
of CPUs, GPUs, DSPs, and other devices~\cite{Stone2010}. However,
achieving portable performance of OpenCL programs is a hard task ---
OpenCL kernels are sensitive to properties of the underlying hardware,
to the implementation, and even to the \emph{dataset} that is operated
upon. This forces developers to laboriously hand tune performance on a
case-by-case basis, since simple heuristics fail to exploit the
available performance.

In this paper, we demonstrate how machine learning-enabled autotuning
can address this issue for one such optimisation parameter of OpenCL
programs --- that of workgroup size. The 2D optimisation space of
OpenCL kernel workgroup sizes is complex and non-linear, making it
resistant to analytical modelling. Successfully applying machine
learning to such a space requires plentiful training data, the careful
selection of features, and an appropriate classification approach. The
approaches presented in this paper use features extracted from the
architecture and kernel, and training data collected from synthetic
benchmarks to predict workgroup sizes for unseen programs.

\section{The SkelCL Stencil Pattern}

\begin{figure}
\centering
\includegraphics[width=.75\columnwidth]{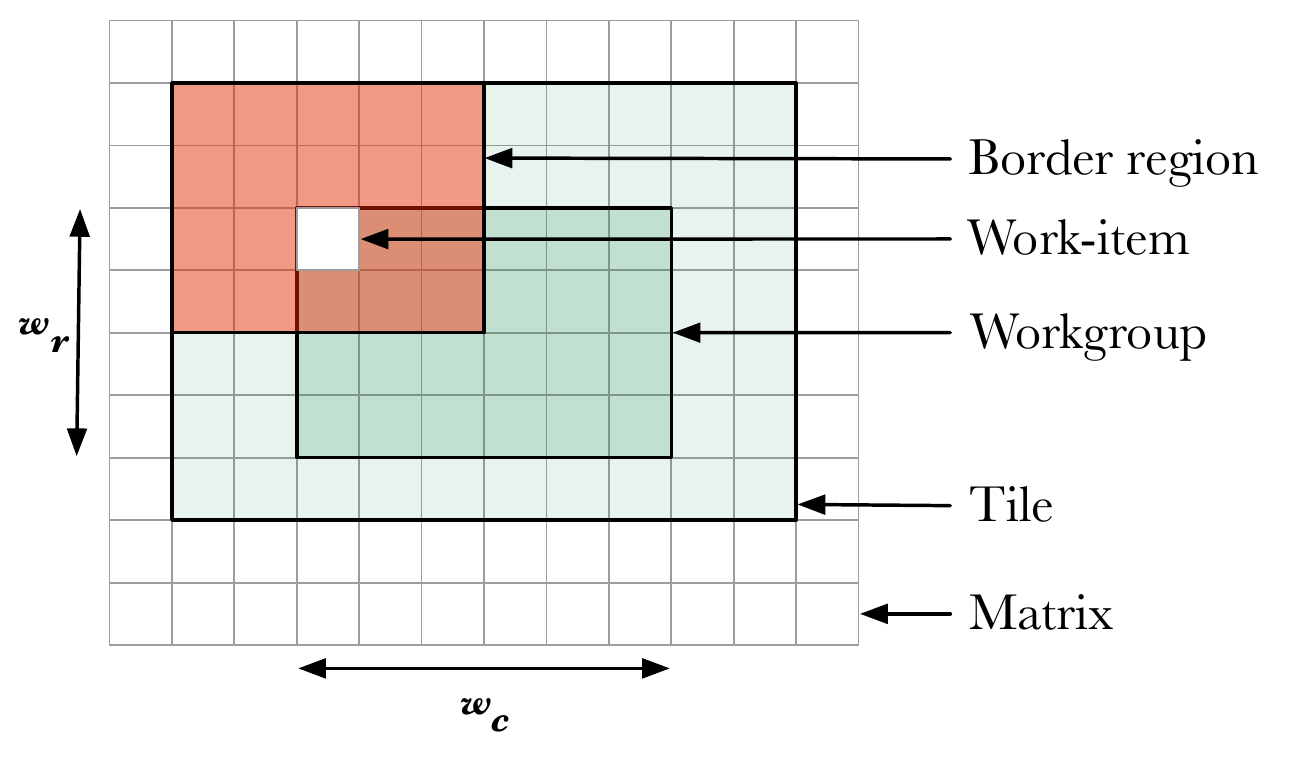}
\caption[Stencil border region]{%
  The components of a stencil: an input matrix is decomposed into
  workgroups, consisting of $w_r \times w_c$ elements. Each element is
  mapped to a work-item. Each work-item operates on its corresponding
  element and a surrounding border region (in this example, 1 element
  to the south, and 2 elements to the north, east, and west).
  \vspace{-1em}
}
\label{fig:stencil-shape}
\end{figure}

Introduced in~\cite{Steuwer2011}, SkelCL is an Algorithmic Skeleton
library which provides OpenCL implementations of data parallel
patterns for heterogeneous parallelism using CPUs and
multi-GPUs. Figure~\ref{fig:stencil-shape} shows the components of the
SkelCL stencil pattern, which applies a user-provided
\emph{customising function} to each element of a 2D matrix. The value
of each element is updated based on its current value and the value of
one or more neighbouring elements, called the \emph{border
  region}. The border region describes a rectangular region about each
cell, and is defined in terms of the number of cells in the border
region to the north, east, south, and west of each cell. Where
elements of a border region fall outside of the matrix bounds, values
are substituted from either a predefined padding value, or the value
of the nearest cell within the matrix, determined by the user.

When a SkelCL stencil pattern is executed, each of the matrix elements
are mapped to OpenCL work-items; and this collection of work-items is
divided into \emph{workgroups} for execution on the target hardware. A
work-item reads the value of its corresponding matrix element and the
surrounding elements defined by the border region. Since the border
regions of neighbouring elements overlap, each element in the matrix
is read multiple times. Because of this, a \emph{tile} of elements of
the size of the workgroup and the perimeter border region is allocated
as a contiguous block in local memory. This greatly reduces the
latency of repeated memory accesses performed by the work-items. As a
result, changing the workgroup size affects both the number of
workgroups which can be active simultaneously, and the amount of local
memory required for each workgroup. While the user defines the size,
type, and border region of the matrix being operated upon, it is the
responsibility of the SkelCL stencil implementation to select an
appropriate workgroup size to use.

\section{Autotuning Workgroup Size}

Selecting the appropriate workgroup size for an OpenCL kernel depends
on the properties of the kernel itself, underlying architecture, and
dataset. For a given \emph{scenario} (that is, a combination of
kernel, architecture, and dataset), the goal of this work is to
harness machine learning to \emph{predict} a performant workgroup size
to use, based on some prior knowledge of the performance of workgroup
sizes for other scenarios. In this section, we describe the
optimisation space and the steps required to apply machine
learning. The autotuning algorithms are described in
Section~\ref{sec:ml}.

\subsection{Constraints}

The space of possible workgroup sizes $W$ is constrained by properties
of both the architecture and kernel. Each OpenCL device imposes a
maximum workgroup size which can be statically checked through the
OpenCL Device API. This constraint reflects architectural limitations
of how code is mapped to the underlying execution hardware. Typical
values are powers of two, e.g.\ 1024, 4096, 8192. Additionally, the
OpenCL runtime enforces a maximum workgroup size on a per-kernel
basis. This value can be queried at runtime once a program has been
compiled for a specific execution device. Factors which affect a
kernel's maximum workgroup size include the number of registers
required, and the available number of SIMD execution units for each
type of executable instruction.

While in theory, any workgroup size which satisfies the device and
kernel workgroup size constraints should provide a valid program, in
practice we find that some combinations of scenario and workgroup size
cause a \texttt{CL\_OUT\_OF\_RESOURCES} error to be thrown when the
kernel is launched. We refer to these workgroup sizes as \emph{refused
  parameters}. Note that in many OpenCL implementations, this error
type acts as a generic placeholder and may not necessarily indicate
that the underlying cause of the error was due to finite resources
constraints. We define the space of \emph{legal} workgroup sizes for a
given scenario $s$ as those which satisfy the architectural and kernel
constraints, and are not refused:
\begin{equation}
  \footnotesize
  W_{legal}(s) = \left\{w | w \in W, w < W_{\max}(s) \right\} - W_{refused}(s)
\end{equation}
Where $W_{\max}(s)$ can be determined at runtime prior to the kernels
execution, but the set $W_{refused}(s)$ can only be discovered
emergently. The set of \emph{safe} parameters are those which are
legal for all scenarios:
\begin{equation}
  % \footnotesize
  W_{safe} = \cap \left\{ W_{legal}(s) | s \in S \right\}
\end{equation}

\subsection{Stencil and Architectural Features}

Since properties of the architecture, program, and dataset all
contribute to the performance of a workgroup size for a particular
scenario, the success of a machine learning system depends on the
ability to translate these properties into meaningful explanatory
variables --- \emph{features}. For each scenario, 102 features are
extracted describing the architecture, kernel, and dataset.

Architecture features are extracted using the OpenCL Device API to
query properties such as the size of local memory, maximum work group
size, and number of compute units. Kernel features are extracted from
the source code stencil kernels by compiling first to LLVM IR bitcode,
and using statistics passes to obtain static instruction counts for
each type of instruction present in the kernel, as well as the total
number of instructions. These instruction counts are divided by the
total number of instructions to produce instruction
\emph{densities}. Dataset features include the input and output data
types, and the 2D matrix dimensions.

\subsection{Training Data}\label{subsec:training}

Training data is collected by measuring the runtimes of stencil
programs using different workgroup sizes. These stencil programs are
generated synthetically using a statistical template substitution
engine, which allows a larger exploration of the program space than is
possible using solely hand-written benchmarks. A stencil template is
parameterised first by stencil shape (one parameter for each of the
four directions), input and output data types (either integers, or
single or double precision floating points), and \emph{complexity} ---
a simple boolean metric for indicating the desired number of memory
accesses and instructions per iteration, reflecting the relatively
bi-modal nature of stencil codes, either compute intensive (e.g.\
finite difference time domain and other PDE solvers), or lightweight
(e.g.\ Game of Life and Gaussian blur).

\section{Machine Learning Methods}\label{sec:ml}

The aim of this work is to design a system which predicts performant
workgroup sizes for \emph{unseen} scenarios, given a set of prior
performance observations. This section presents three contrasting
methods for achieving this goal.

\subsection{Predicting Oracle Workgroup Sizes}

\begin{algorithm}[t]
\begin{algorithmic}[1]
\Require scenario $s$
\Ensure workgroup size $w$

\Procedure{Baseline}{s}
\State $w \leftarrow \text{classify}(f(s))$
\If{$w \in W_{legal}(s)$}
    \State \textbf{return} $w$
\Else
  \State \textbf{return} $\underset{w \in W_{safe}}{\argmax}
\left(
  \prod_{s \in S_{training}} p(s, w)
\right)^{1/|S_{training}|}$
\EndIf
\EndProcedure
\item[] % line break

\Procedure{Random}{s}
\State $w \leftarrow \text{classify}(f(s))$
\While{$w \not\in W_{legal}(s)$}
  \State $W \leftarrow \left\{ w | w < W_{max}(s), w \not\in W_{refused}(s) \right\}$
  \State $w \leftarrow $ random selection $w \in W$
\EndWhile
\State \textbf{return} $w$
\EndProcedure
\item[] % line break

\Procedure{NearestNeighbour}{s}
\State $w \leftarrow \text{classify}(f(s))$
\While{$w \not\in W_{legal}(s)$}
  \State $d_{min} \leftarrow \infty$
  \State $w_{closest} \leftarrow \text{null}$
  \For{$c \in \left\{ w | w < W_{\max}(s), w \not\in W_{refused}(s) \right\}$}
    \State $d \leftarrow \sqrt{\left(c_r - w_r\right)^2 + \left(c_c - w_c\right)^2}$
    \If{$d < d_{min}$}
      \State $d_{min} \leftarrow d$
      \State $w_{closest} \leftarrow c$
    \EndIf
  \EndFor
  \State $w \leftarrow w_{closest}$
\EndWhile
\State \textbf{return} $w$
\EndProcedure
\end{algorithmic}
\caption{Prediction using classifiers}
\label{alg:autotune-classification}
\end{algorithm}

The first approach is detailed in
Algorithm~\ref{alg:autotune-classification}. By considering the set of
possible workgroup sizes as a hypothesis space, we train a classifier
to predict, for a given set of features, the \emph{oracle} workgroup
size. The oracle workgroup size $\Omega(s)$ is the workgroup size
which provides the lowest mean runtime $t(s,w)$ for a scenario $s$:
\begin{equation}
  \Omega(s) = \argmin_{w \in W_{legal}(s)} t(s,w)
\end{equation}
Training a classifier for this purpose requires pairs of stencil
features $f(s)$ to be labelled with their oracle workgroup size for a
set of training scenarios $S_{training}$:
\begin{equation}
  D_{training} = \left\{ \left(f(s), \Omega(s)\right) | s \in S_{training} \right\}
\end{equation}
After training, the classifier predicts workgroup sizes for unseen
scenarios from the set of oracle workgroup sizes from the training
set. This is a common and intuitive approach to autotuning, in that a
classifier predicts the best parameter value based on what worked well
for the training data. However, given the constrained space of
workgroup sizes, this presents the problem that future scenarios may
have different sets of legal workgroup sizes to that of the training
data, i.e.:
\begin{equation}
  \bigcup_{\forall s \in S_{future}} W_{legal}(s) \nsubseteq \left\{ \Omega(s) | s \in S_{training} \right\}
\end{equation}
This results in an autotuner which may predict workgroup sizes that
are not legal for all scenarios, either because they exceed
$W_{\max}(s)$, or because parameters are refused,
$w \in W_{refused}(s)$. For these cases, we evaluate the effectiveness
of three \emph{fallback handlers}, which will iteratively select new
workgroup sizes until a legal one is found:
\begin{enumerate}
\item \emph{Baseline} --- select the workgroup size which provides the
  highest average case performance from the set of safe workgroup sizes.
\item \emph{Random} --- select a random workgroup size which is
  expected from prior observations to be legal.
\item \emph{Nearest Neighbour} --- select the workgroup size which
  from prior observations is expected to be legal, and has the lowest
  Euclidian distance to the prediction.
\end{enumerate}

\subsection{Predicting Kernel Runtimes}

\begin{algorithm}[t]
\begin{algorithmic}[1]
\Require scenario $s$, regressor $R(x, w)$, fitness function $\Delta(x)$
\Ensure workgroup size $w$

\State $W \leftarrow \left\{ w | w < W_{\max}(s) \right\} -
W_{refused}(s)$
\Comment Candidates.
\State $w \leftarrow \underset{w \in W}{\argmax} \Delta(R(f(s), w))$
\Comment Select best candidate.
\While{$w \not\in W_{legal}(s)$}
  \State $W_{refused}(s) = W_{refused}(s) + \{w\}$
  \State $W \leftarrow W - \left\{ w \right\}$
  \Comment Remove candidate from selection.
  \State $w \leftarrow \underset{w \in W}{\argmax} \Delta(R(f(s), w))$
  \Comment Select best candidate.
\EndWhile
\State \textbf{return} $w$
\end{algorithmic}
\caption{Prediction using regressors}
\label{alg:autotune-regression}
\end{algorithm}

A problem of predicting oracle workgroup sizes is that, for each
training instance, an exhaustive search of the optimisation space must
be performed in order to find the oracle workgroup size. An
alternative approach is to instead predict the expected \emph{runtime}
of a kernel given a specific workgroup size. Given training data
consisting of $(f(s),w,t)$ tuples, where $f(s)$ are scenario features,
$w$ is the workgroup size, and $t$ is the observed runtime, we train a
regressor $R(f(s), w)$ to predict the runtime of scenario and
workgroup size combinations. The selected workgroup size
$\bar{\Omega}(s)$ is then the workgroup size from a pool of candidates
which minimises the output of the
regressor. Algorithm~\ref{alg:autotune-regression} formalises this
approach of autotuning with regressors. A fitness function $\Delta(x)$
computes the reciprocal of the predicted runtime so as to favour
shorter over longer runtimes. Note that the algorithm is self
correcting in the presence of refused parameters --- if a workgroup
size is refused, it is removed from the candidate pool, and the next
best candidate is chosen. This removes the need for fallback
handlers. Importantly, this technique allows for training on data for
which the oracle workgroup size is unknown, meaning that a full
exploration of the space is not required in order to gather a training
instance, as is the case with classifiers.

\subsection{Predicting Relative Performance}

Accurately predicting the runtime of arbitrary code is a difficult
problem. It may instead be more effective to predict the relative
performance of two different workgroup sizes for the same kernel. To
do this, we predict the \emph{speedup} of a workgroup size over a
baseline. This baseline is the workgroup size which provides the best
average case performance across all scenarios and is known to be
safe. Such a baseline value represents the \emph{best} possible
performance which can be achieved using a single, fixed workgroup
size. As when predicting runtimes, this approach performs
classification using regressors
(Algorithm~\ref{alg:autotune-regression}). We train a regressor
$R(f(s), w)$ to predict the relative performance of workgroup size $w$
over a baseline parameter for scenario $s$. The fitness function
returns the output of the regressor, so the selected workgroup size
$\bar{\Omega}(s)$ is the workgroup size from a pool of candidates
which is predicted to provide the best relative performance. This has
the same advantageous properties as predicting runtimes, but by
training using relative performance, we negate the challenges of
predicting dynamic code behaviour.

\section{Experimental Setup}

\begin{table*}
\scriptsize
\centering
\begin{tabular}{l r | l r r r r r}
\toprule
Host & Host Memory &  OpenCL Device &  Compute units & Frequency & Local Memory & Global Cache & Global Memory \\
\midrule
Intel i5-2430M & 8 GB  & CPU              &              4 &   2400 Hz &        32 KB &       256 KB &       7937 MB \\
Intel i5-4570  & 8 GB  & CPU              &              4 &   3200 Hz &        32 KB &       256 KB &       7901 MB \\
Intel i7-3820  & 8 GB  & CPU              &              8 &   1200 Hz &        32 KB &       256 KB &       7944 MB \\
Intel i7-3820  & 8 GB  & AMD Tahiti 7970  &             32 &   1000 Hz &        32 KB &        16 KB &       2959 MB \\
Intel i7-3820  & 8 GB  & Nvidia GTX 590   &              1 &   1215 Hz &        48 KB &       256 KB &       1536 MB \\
Intel i7-2600K & 16 GB & Nvidia GTX 690   &              8 &   1019 Hz &        48 KB &       128 KB &       2048 MB \\
Intel i7-2600  & 8 GB  & Nvidia GTX TITAN &             14 &    980 Hz &        48 KB &       224 KB &       6144 MB \\
\bottomrule
\end{tabular}
\caption{Specification of experimental platforms and OpenCL devices.}
\label{tab:hw}
\end{table*}

To evaluate the performance of the presented autotuning techniques, an
exhaustive enumeration of the workgroup size optimisation space for
429 combinations of architecture, program, and dataset was performed.

Table~\ref{tab:hw} describes the experimental platforms and OpenCL
devices used. Each platform was unloaded, frequency governors
disabled, and benchmark processes set to the highest priority
available to the task scheduler. Datasets and programs were stored in
an in-memory file system. All runtimes were recorded with millisecond
precision using OpenCL's Profiling API to record the kernel execution
time. The workgroup size space was enumerated for each combination of
$w_r$ and $w_c$ values in multiples of 2, up to the maximum workgroup
size. For each combination of scenario and workgroup size, a minimum
of 30 runtimes were recorded.

In addition to the synthetic stencil benchmarks described in
Section~\ref{subsec:training}, six stencil kernels taken from four
reference implementations of standard stencil applications from the
fields of image processing, cellular automata, and partial
differential equation solvers are used: Canny Edge Detection, Conway's
Game of Life, Heat Equation, and Gaussian
Blur. Table~\ref{tab:kernels} shows details of the stencil kernels for
these reference applications and the synthetic training benchmarks
used. Dataset sizes of size $512\times512$, $1024\times1024$,
$2048\times2048$, and $4096\times4096$ were used.

Program behavior is validated by comparing program output against a
gold standard output collected by executing each of the real-world
benchmarks programs using the baseline workgroup size. The output of
real-world benchmarks with other workgroup sizes is compared to this
gold standard output to test for correct program execution.

Five different classification algorithms are used to predict oracle
workgroup sizes, chosen for their contrasting properties: Naive Bayes,
SMO, Logistic Regression, J48 Decision tree, and Random
Forest~\cite{Han2011}. For regression, a Random Forest with regression
trees is used, chosen because of its efficient handling of large
feature sets compared to linear models~\cite{Breiman1999}. The
autotuning system is implemented in Python as a system daemon. SkelCL
stencil programs request workgroup sizes from this daemon, which
performs feature extraction and classification.

\section{Performance Results}\label{sec:results}

\begin{table}
\scriptsize
\centering
\begin{tabular}{lrrrrR{1.4cm}}
\toprule
      Name &  North &  South &  East &  West &  Instruction Count \\
\midrule
   synthetic-a & 1--30 & 1--30 & 1--30 & 1--30 & 67--137\\
   synthetic-b & 1--30 & 1--30 & 1--30 & 1--30 & 592--706\\
   gaussian    & 1--10 & 1--10 & 1--10 & 1--10 & 82--83 \\
   gol         &      1 &      1 &     1 &     1 &                190 \\
   he          &      1 &      1 &     1 &     1 &                113 \\
   nms         &      1 &      1 &     1 &     1 &                224 \\
   sobel       &      1 &      1 &     1 &     1 &                246 \\
   threshold   &      0 &      0 &     0 &     0 &                 46 \\
\bottomrule
\end{tabular}
\caption{%
  Stencil kernels, border sizes (north, south, east, and west),
  and static instruction counts.
  \vspace{-1.5em}
}
\label{tab:kernels}
\end{table}

This section describes the performance results of enumerating the
workgroup size optimisation space. The effectiveness of autotuning
techniques for exploiting this space are examined in
Section~\ref{sec:evaluation}. The experimental results consist of
measured runtimes for a set of \emph{test cases}, where a test case
$\tau_i$ consists of a scenario, workgroup size pair
$\tau_i = (s_i,w_i)$, and is associated with a \emph{sample} of
observed runtimes of the program. A total of 269813 test cases were
evaluated, which represents an exhaustive enumeration of the workgroup
size optimisation space for 429 scenarios. For each scenario, runtimes
for an average of 629 (max 7260) unique workgroup sizes were
measured. The average sample size for each test case is 83 (min 33,
total 16917118).

\begin{figure}
  \centering
  \vspace{-2em}
  \includegraphics[width=1.08\columnwidth]{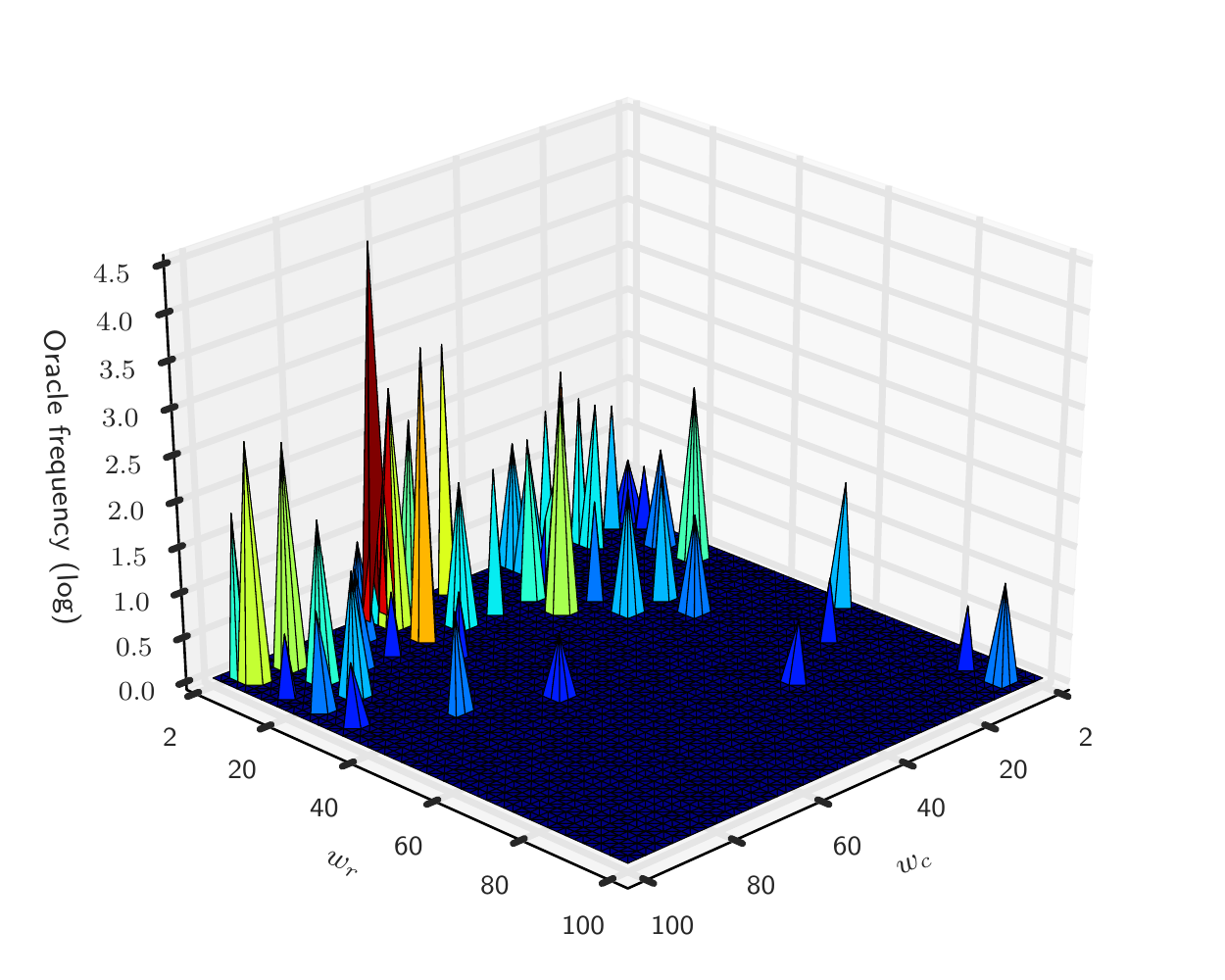}
  \vspace{-2.5em}
  \caption{%
    Oracle frequency counts for a subset of the workgroup sizes,
    $w_c \le 100, w_r \le 100$. There are 135 unique oracle workgroup
    sizes. The most common oracle workgroup size is
    $w_{(64 \times 4)}$, optimal for 15\% of scenarios.
    \vspace{-2em}
  }
\label{fig:oracle-wgsizes}
\end{figure}

The workgroup size optimisation space is non-linear and complex, as
shown in Figure~\ref{fig:oracle-wgsizes}, which plots the distribution
of optimal workgroup sizes. Across the 429 scenarios, there are 135
distinct optimal workgroup sizes (31.5\%). The average speedup of the
oracle workgroup size over the worst workgroup size for each scenario
is $15.14\times$ (min $1.03\times$, max $207.72\times$).

Of the 8504 unique workgroup sizes tested, 11.4\% were refused in one
or more test cases, with an average of 5.5\% test cases leading to
refused parameters. There are certain patterns to the refused
parameters: for example, workgroup sizes which contain $w_c$ and $w_r$
values which are multiples of eight are less frequently refused, since
eight is a common width of SIMD vector
operations~\cite{IntelCorporation2012}. However, a refused parameter
is an obvious inconvenience to the user, as one would expect that any
workgroup size within the specified maximum should generate a working
program, if not a performant one.

\begin{figure}
  \centering
  \vspace{-2em}
  \includegraphics[width=.6\columnwidth]{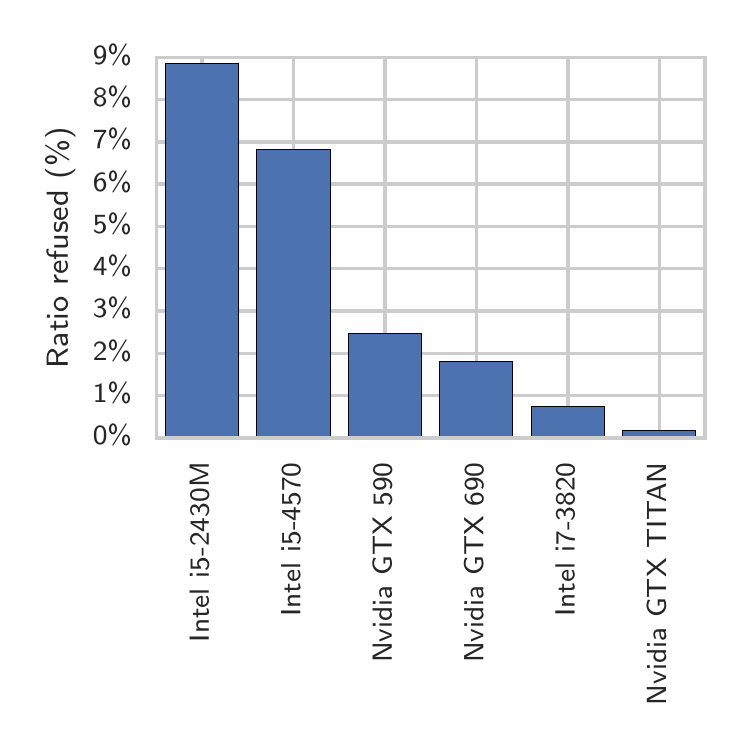}
  \vspace{-1.5em}
  \caption{%
    The ratio of test cases with refused workgroup
    sizes, grouped by OpenCL device ID. No parameters were refused by
    the AMD device.
    \vspace{-1em}
  }
\label{fig:refused-params}
\end{figure}

Experimental results suggest that the problem is vendor --- or at
least device --- specific. Figure~\ref{fig:refused-params} shows the
ratio of refused test cases, grouped by device. We see many more
refused parameters for test cases on Intel CPU devices than any other
type, while no workgroup sizes were refused by the AMD GPU. The exact
underlying cause for these refused parameters is unknown, but can
likely by explained by inconsistencies or errors in specific OpenCL
driver implementations. Note that the ratio of refused parameters
decreases across the three generations of Nvidia GPUs: GTX 590 (2011),
GTX 690 (2012), and GTX TITAN (2013). For now, it is imperative that
any autotuning system is capable of adapting to these refused
parameters by suggesting alternatives when they occur.

The baseline parameter is the workgroup size providing the best
overall performance while being legal for all scenarios. Because of
refused parameters, only a \emph{single} workgroup size
$w_{(4 \times 4)}$ from the set of experimental results is found to
have a legality of 100\%, suggesting that an adaptive approach to
setting workgroup size is necessary not just for the sake of
maximising performance, but also for guaranteeing program
execution. The utility of the baseline parameter is that it represents
the best performance that can be achieved through static tuning of the
workgroup size parameter; however, compared to the oracle workgroup
size for each scenario, the baseline parameter achieves only 24\% of
the optimal performance.

\section{Evaluation of Autotuning Methods}\label{sec:evaluation}

In this section we evaluate the effectiveness of the three proposed
autotuning techniques for predicting performant workgroup sizes. For
each autotuning technique, we partition the experimental data into
training and testing sets. Three strategies for partitioning the data
are used: the first is a 10-fold cross-validation; the second is to
divide the data such that only data collected from synthetic
benchmarks are used for training and only data collected from the
real-world benchmarks are used for testing; the third strategy is to
create leave-one-out partitions for each unique device, kernel, and
dataset. For each combination of autotuning technique and testing
dataset, we evaluate each of the workgroup sizes predicted for the
testing data using the following metrics:
\begin{itemize}
\item time (real) --- the time taken to make the autotuning
  prediction. This includes classification time and any communication
  overheads.
\item accuracy (binary) --- whether the predicted workgroup size is
  the true oracle, $w = \Omega(s)$.
\item validity (binary) --- whether the predicted workgroup size
  satisfies the workgroup size constraints constraints,
  $w < W_{\max}(s)$.
\item refused (binary) --- whether the predicted workgroup size is
  refused, $w \in W_{refused}(s)$.
\item performance (real) --- the performance of the predicted
  workgroup size relative to the oracle for that scenario.
\item speedup (real) --- the relative performance of the predicted
  workgroup size relative to the baseline workgroup size
  $w_{(4 \times 4)}$.
\end{itemize}
The \emph{validty} and \emph{refused} metrics measure how often
fallback strategies are required to select a legal workgroup size
$w \in W_{legal}(s)$. This is only required for the classification
approach to autotuning, since the process of selecting workgroup sizes
using regressors respects workgroup size constraints.

\subsection{Predicting Oracle Workgroup Size}

Figure~\ref{fig:class-syn} shows the results when classifiers are
trained using data from synthetic benchmarks and tested using
real-world benchmarks. With the exception of the ZeroR, a dummy
classifier which ``predicts'' only the baseline workgroup size
$w_{\left( 4 \times 4 \right)}$, the other classifiers achieve good
speedups over the baseline, ranging from $4.61\times$ to $5.05\times$
when averaged across all test sets. The differences in speedups
between classifiers is not significant, with the exception of
SimpleLogistic, which performs poorly when trained with synthetic
benchmarks and tested against real-world programs. This suggests the
model over-fitting to features of the synthetic benchmarks which are
not shared by the real-world tests. Of the three fallback handlers,
\textsc{NearestNeighbour} provides the best performance, indicating
that it successfully exploits structure in the optimisation space. In
our evaluation, the largest number of iterations of a fallback handler
required before selecting a legal workgroup size was 2.

\subsection{Predicting with Regressors}

Figure~\ref{fig:regression-class} shows a summary of results for
autotuning using regressors to predict kernel runtimes
(\ref{fig:runtime-class-xval}) and speedups
(\ref{fig:speedup-class-xval}). Of the two regression techniques,
predicting the \emph{speedup} of workgroup sizes is much more
successful than predicting the \emph{runtime}. This is most likely
caused by the inherent difficulty in predicting the runtime of
arbitrary code, where dynamic factors such as flow control and loop
bounds are not captured by the instruction counts which are used as
features for the machine learning models. The average speedup achieved
by predicting runtimes is $4.14\times$. For predicting speedups, the
average is $5.57\times$, the highest of all of the autotuning
techniques.

\subsection{Autotuning Overheads}

\begin{figure}
\centering
\includegraphics[width=\columnwidth]{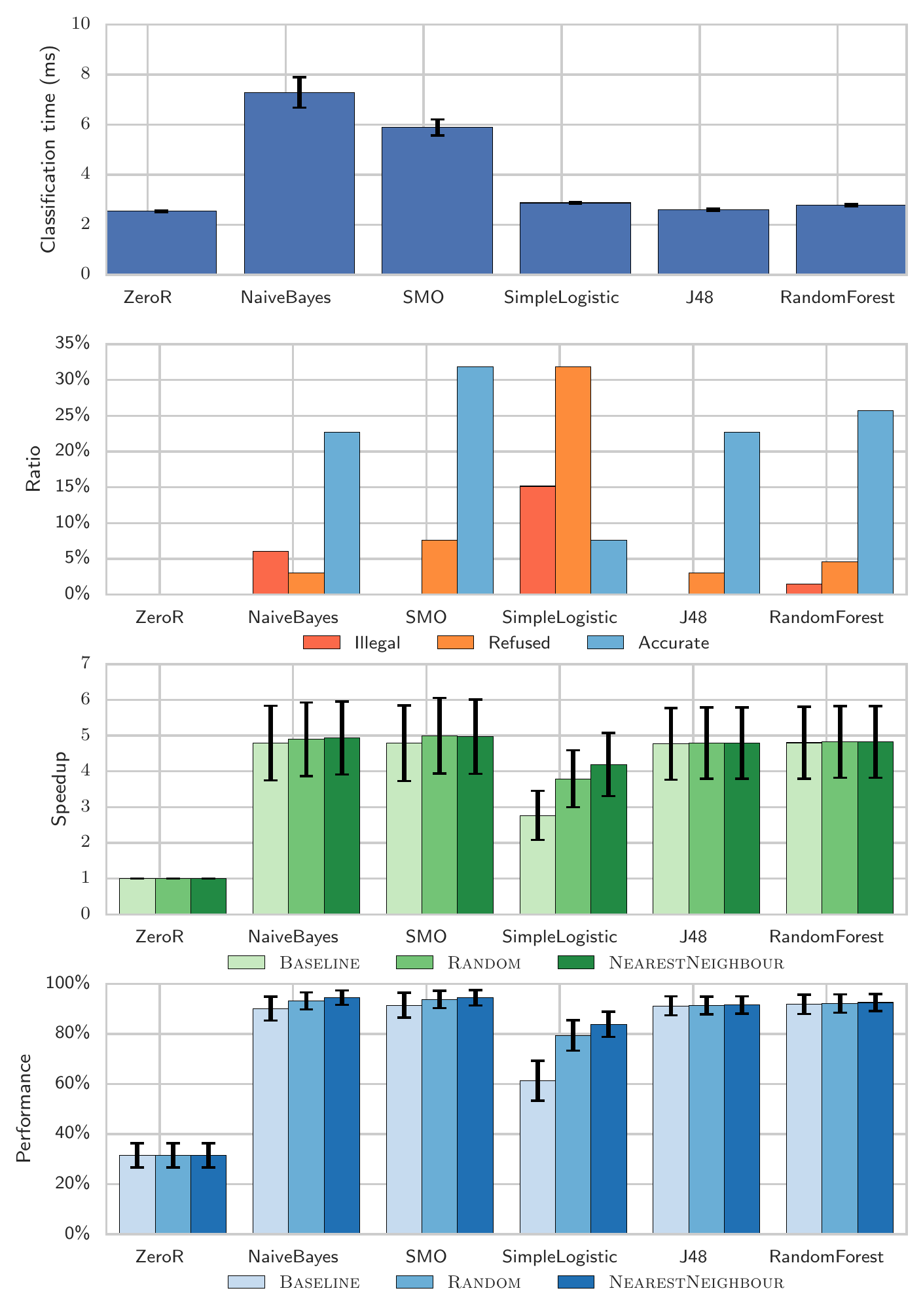}
\vspace{-2em}
\caption{%
  Autotuning performance using classifiers and synthetic
  benchmarks. Each classifier is trained on data collected from
  synthetic stencil applications, and tested for prediction quality
  using data from 6 real-world benchmarks. 95\% confidence intervals
  are shown where appropriate.
  \vspace{-1.5em}%
}
\label{fig:class-syn}
\end{figure}

\begin{figure}
\centering
\begin{subfigure}[h]{.48\columnwidth}
\centering
\includegraphics[width=\columnwidth]{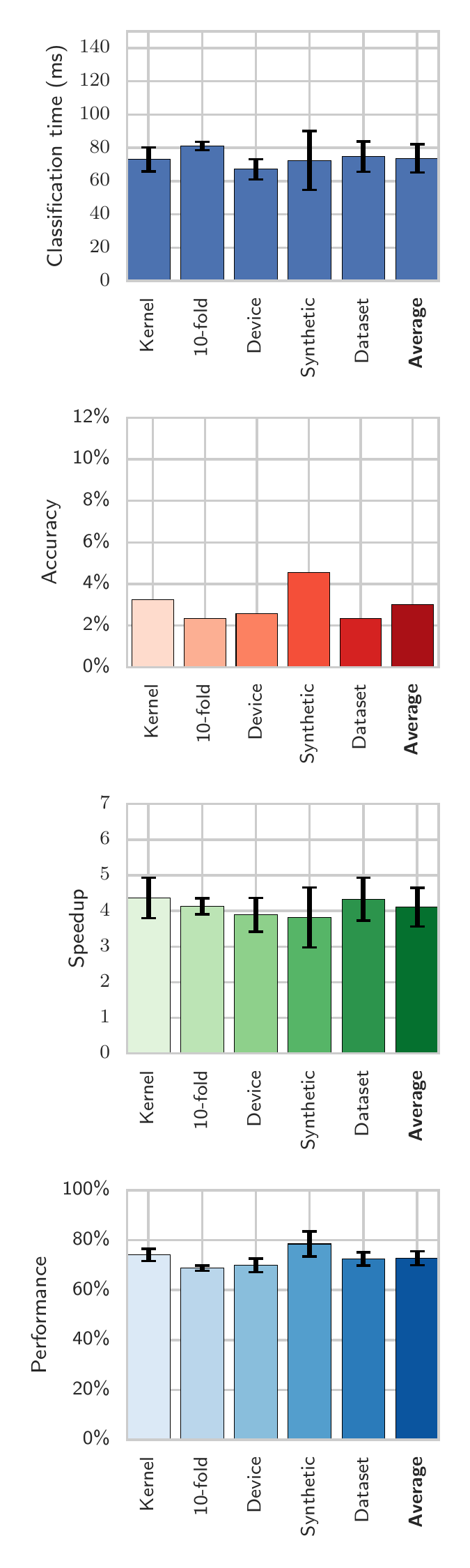}
\vspace{-2em}
\caption{}
\label{fig:runtime-class-xval}
\end{subfigure}
\begin{subfigure}[h]{.48\columnwidth}
\centering
\includegraphics[width=\columnwidth]{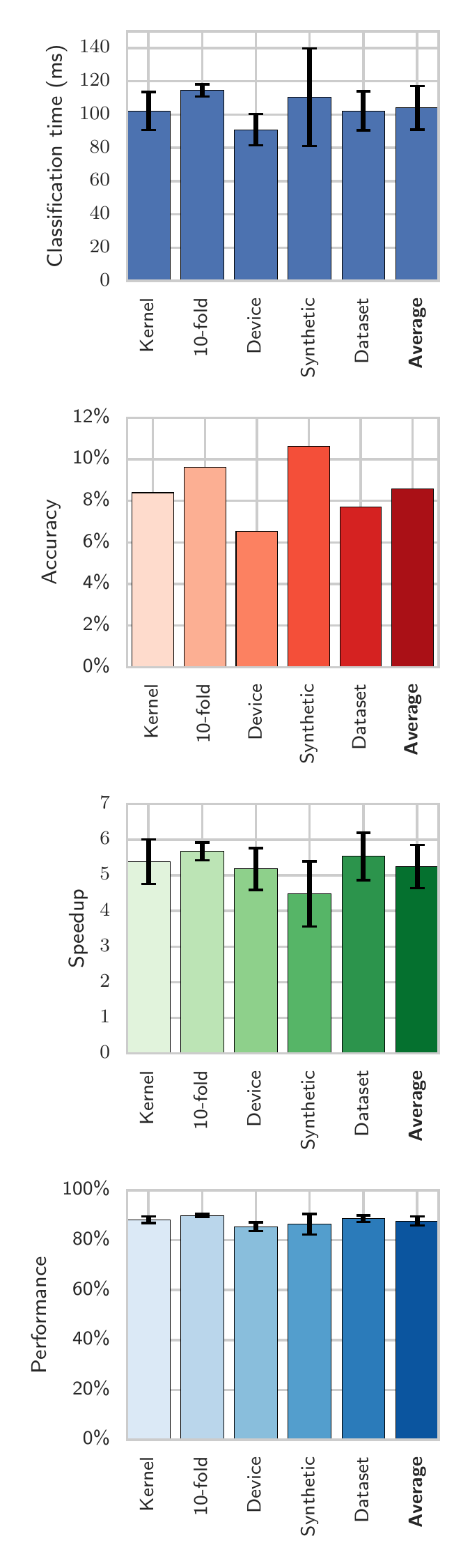}
\vspace{-2em}
\caption{}
\label{fig:speedup-class-xval}
\end{subfigure}
\vspace{-.5em}
\caption{%
  Autotuning performance for each type of test dataset using
  regressors to predict: (\subref{fig:runtime-class-xval}) kernel
  runtimes, and (\subref{fig:speedup-class-xval}) relative performance
  of workgroup sizes.
  \vspace{-1em} %
}
\label{fig:regression-class}
\end{figure}

Comparing the classification times of Figures~\ref{fig:class-syn}
and~\ref{fig:regression-class} shows that the prediction overhead of
regressors is significantly greater than classifiers. This is because,
while a classifier makes a single prediction, the number of
predictions required of a regressor grows with the size of
$W_{\max}(s)$, since classification with regression requires making
predictions for all $w \in \left\{ w | w < W_{\max}(s) \right\}$. The
fastest classifier is J48, due to the it's simplicity --- it can be
implemented as a sequence of nested \texttt{if} and \texttt{else}
statements.

\subsection{Comparison with Human Expert}

\begin{figure}
\centering
\includegraphics[width=\columnwidth]{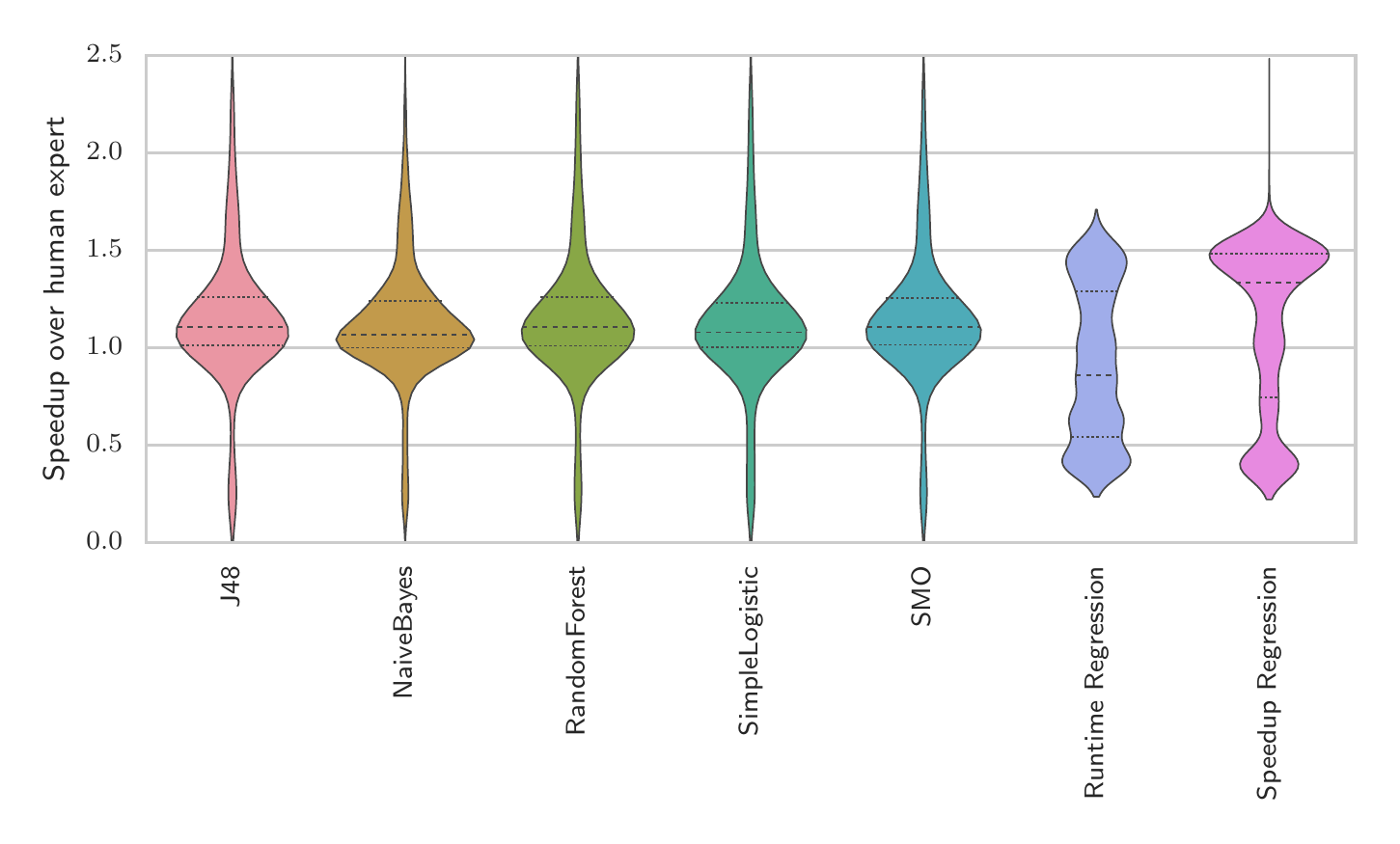}
\vspace{-2.5em}
\caption[Speedup results over human expert]{%
  Violin plot of speedups over \emph{human expert}, ignoring cases
  where the workgroup size selected by human experts is
  invalid. Classifiers are using \textsc{NearestNeighbour} fallback
  handlers. Horizontal dashed lines show the median, Q1, and
  Q3. Kernel Density Estimates show the distribution of results. The
  speedup axis is fixed to the range 0--2.5 to highlight the IQRs,
  which results in some outlier speedups > 2.5 being clipped.
  \vspace{-.8em} %
}
\label{fig:speedup-distributions}
\end{figure}

In the original implementation of the SkelCL stencil
pattern~\cite{Steuwer2014a}, Steuwer et al.\ selected a workgroup size
of $w_{(32 \times 4)}$ in an evaluation of 4 stencil operations on a
Tesla S1070 system. In our evaluation of 429 combinations of kernel,
architecture, and dataset, we found that this workgroup size is
refused by 2.6\% of scenarios, making it unsuitable for use as a
baseline. However, if we remove the scenarios for which
$w_{(32 \times 4)}$ is \emph{not} a legal workgroup size, we can
directly compare the performance against the autotuning predictions.

Figure~\ref{fig:speedup-distributions} plots the distributions and
Interquartile Range (IQR) of all speedups over the human expert
parameter for each autotuning technique. The distributions show
consistent classification results for the five classification
techniques, with the speedup at Q1 for all classifiers being
$\ge 1.0\times$. The IQR for all classifiers is $< 0.5$, but there are
outliers with speedups both well below $1.0\times$ and well above
$2.0\times$. In contrast, the speedups achieved using regressors to
predict runtimes have a lower range, but also a lower median and a
larger IQR. Clearly, this approach is the least effective of the
evaluated autotuning techniques. Using regressors to predict relative
performance is more successful, achieving the highest median speedup
of all the techniques ($1.33\times$).

\section{Related Work}\label{sec:related}

Ganapathi et al.\ demonstrated early attempts at autotuning multicore
stencil codes in~\cite{Ganapathi2009}, drawing upon the successes of
statistical machine learning techniques in the compiler
community. They use Kernel Canonical Correlation Analysis to build
correlations between stencil features and optimisation
parameters. Their use of KCCA restricts the scalability of their
system, as the complexity of model building grows exponentially with
the number of features. A code generator and autotuner for 3D Jacobi
stencil codes is presented in~\cite{Zhang2013a}, although their
approach requires a full enumeration of the parameter space for each
new program, and has no cross-program learning. Similarly,
CLTune~\cite{Nugteren2015} is an autotuner which applies iterative
search techniques to user-specified OpenCL parameters. The number of
parallel mappers and reducers for MapReduce workloads is tuned
in~\cite{Johnston2015a} using surrogate models rather than machine
learning, although the optimisation space is not subject to the level
of constraints that OpenCL workgroup size is. A generic OpenCL
autotuner is presented in~\cite{Falch2015} which uses neural networks
to predict good configurations of user-specified parameters, although
the authors present only a preliminary evaluation using three
benchmarks. Both systems require the user to specify parameters on a
per-program basis. The autotuner presented in this work, embedded at
the skeletal level, requires no user effort for new programs and is
transparent to the user. A DSL and CUDA code generator for stencils is
presented in~\cite{Kamil2010}. Unlike the SkelCL stencil pattern, the
generated stencil codes do not exploit fast local device memory. The
automatic generation of synthetic benchmarks using parameterised
template substitution is presented in~\cite{Chiu2015}. The authors
describe an application of their tool for generating OpenCL stencil
kernels for machine learning, but do not report any performance
results.

\section{Conclusions}\label{sec:conclusions}

We present and compare novel methodologies for autotuning the
workgroup size of stencil patterns using the established open source
library SkelCL. These techniques achieve up to 94\% of the maximum
performance, while providing robust fallbacks in the presence of
unexpected behaviour in OpenCL driver implementations. Of the three
techniques proposed, predicting the relative performances of workgroup
sizes using regressors provides the highest median speedup, whilst
predicting the oracle workgroup size using decision tree classifiers
adds the lowest runtime overhead. This presents a trade-off between
classification time and training time that could be explored in future
work using a hybrid of the classifier and regressor techniques
presented in this paper.

In future work, we will extend the autotuner to accommodate additional
OpenCL optimisation parameters and skeleton patterns. Feature
selection can be evaluated using Principle Component Analysis, as well
exploring the relationship between prediction accuracy and the number
of synthetic benchmarks used. A promising avenue for further research
is in the transition towards online machine learning which is enabled
by using regressors to predict kernel runtimes. This could be combined
with the use of adaptive sampling plans to minimise the number of
observations required to distinguish bad from good parameter values,
such as presented in~\cite{Leather2009}. Dynamic profiling can be
used to increase the prediction accuracy of kernel runtimes by
capturing the runtime behaviour of stencil kernels.

\acks

This work was supported by the UK Engineering and Physical Sciences
Research Council under grants EP/L01503X/1 for the University of
Edinburgh School of Informatics Centre for Doctoral Training in
Pervasive Parallelism
(\url{http://pervasiveparallelism.inf.ed.ac.uk/}), EP/H044752/1
(ALEA), and EP/M015793/1 (DIVIDEND).

\label{bibliography}
\printbibliography

\end{document}